\documentclass[aps,prl,twocolumn,twoside,floatfix,a4paper,superscriptaddress,showpacs]{revtex4}
\usepackage{hyperref}
\usepackage{bbm}
\usepackage{amsmath, amssymb}
\usepackage{color}
\usepackage{graphicx,epsfig}
\usepackage{pifont} 

\def\CZ{C\!Z} 

\def\KO{K} % symbol for Krausoperator
\def\tr{\mbox{tr}}

\newtheorem{prop}{Proposition}\def\PRO{\begin{prop}}\def\ORP{\end{prop}}
\def\TH{\begin{theo}}\def\HT{\end{theo}}
\def\LE{\begin{lemme}}\def\EL{\end{lemme}}

\def\ket#1{| #1\rangle}
\def\bra#1{\langle#1{|}}
\def\<{\langle}
\def\>{\rangle}

\def\ta{\theta}
\def\mop{} % if needed \tilde can be included 

\def\paper{paper }

\begin{document}

\title{Ancilla-Driven Universal Quantum Computation}

\author{Janet Anders}
\affiliation{Department of Physics \& Astronomy, University College London,
London WC1E 6BT, UK}
\author{Daniel K. L. Oi}
\affiliation{SUPA, Department of Physics, University of Strathclyde,
Glasgow G4 0NG, UK}
\author{Elham Kashefi}
\affiliation{School of Informatics, University of Edinburgh,
Edinburgh, EH8 9AB, UK}
\author{Dan E. Browne}
\affiliation{Department of Physics \& Astronomy, University College London,
London WC1E 6BT, UK}
\author{Erika Andersson}
\affiliation{SUPA, Department of Physics, Heriot-Watt University,
Edinburgh EH14 4AS, UK}

\begin{abstract}

We propose a method of manipulating a quantum register remotely with the help of a single ancilla that ``steers'' the evolution of the register. The fully controlled ancilla qubit is coupled to the computational register solely via a \emph{fixed} unitary two-qubit interaction, $E$, and then measured in suitable bases. We characterize all interactions $E$ that induce a unitary, step-wise deterministic measurement back-action on the register sufficient to implement any arbitrary quantum channel. Our scheme offers  significant experimental advantages for implementing computations, preparing states and performing generalized measurements as no direct control of the register is required.

\end{abstract}

\maketitle

%%%
{\it Introduction.--}
The two best-known strategies of performing a quantum computation are the gate-based and measurement-based model (MBQC). In the former, a computation is performed by actively manipulating individual register qubits by a network of logical gates. The required control of the register is very challenging to realize in experiments. MBQC is an alternative strategy that relies on the powerful effect of measurements on entangled quantum systems  \cite{RB01,BB06}. A computation is here implemented ``passively'', as a sequence of adaptive single-qubit measurements on an entangled multi-partite resource state and realized in experiments  \cite{experimental-realisations} were single qubit measurements are ``cheap''. Hard work, however, goes into the construction of the initial resource state.

%  that actively manipulate them. % to produce the register's output state. 

In this \paper we introduce a hybrid model that is tailored to fit many experimental settings. The scheme utilizes a single fully controlled ancilla qubit, which is coupled sequentially to one or, at most, two qubits of a register via a \emph{fixed} entangling operation, $E$, to enable universal state preparation \cite{USP}. After each coupling the ancilla is measured in a suitable basis, providing a back-action onto the register which implements unitary evolution of the register. Moreover, using an additional, second ancilla qubit any Positive Operator Valued Measurement (POVM), and thus any quantum channel, can be realized. The computation requires no direct control of the register nor the preparation of a large entangled state. The processing of information is driven by active manipulation of the ancilla alone and we shall call the model \emph{Ancilla-Driven Quantum Computation} (ADQC).

%\medskip

Previous attempts to construct `remotely controlled', deterministic and universal quantum state preparators have concluded this to be impossible \cite{nielsen,huelga}. These schemes use, like ours, additional `programmable qubits', i.e. ancillas, that are coupled to the register with a fixed interaction. Our results highlight that the existing no-go theorems do not apply when \emph{feedback} within the programmable part  is allowed and a final \emph{local} redefinition of the computational basis of the register is performed. 

%%%%%%%%%%%%%%%%%%
\begin{figure}[b]
   \begin{center}
	\includegraphics[width=0.3\textwidth]{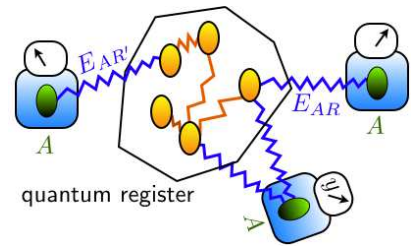}
        \caption{\label{fig:ADQC} Illustration of an ancilla-driven computation
          on a register consisting of several qubits. A single ancilla, $A$, is
          sequentially coupled to one, or at most two, register qubits, $R$ and
          $R'$ etc., and measured. The coupling, $E_{AR}$, is fixed throughout
          the computation while the measurements on the ancilla, indicated by
          the arrows, can differ. }
   \end{center}
\end{figure}
%%%%%%%%%%%%%%%%%%

%\medskip

The key advantage of ADQC is that register qubits are only addressed with a single coupling operation, $E$. No further operation, neither unitary nor measurement, is required. 
%In particular, the register qubits need not ever interact directly but can be entangled with the help of the ancilla. 
This provides architectural advantages for many experimental situations as the computational register can be separated from state preparation and measurement, and does not require bespoke control. This lends itself to systems where long-lived but static qubits are addressed by more mobile ancilla qubits subject to a single entangling interaction. There are many physical systems which possess the required features and interactions, such as neutral atoms in optical lattices \cite{JBCGZ1999}, micro ion trap arrays~\cite{CZ2000}, nuclear-electron spin systems~\cite{Kane1998} and cavity QED-superconducting qubits~\cite{MCGKJSFSHWBDGS2007}. ADQC is also advantageous in physical systems where measurements are destructive, i.e. experiments with photons, as the measurement never acts on the register and leaves it intact for further manipulation.  

\smallskip

{\it Example: Control-Z \& Hadamard interaction. --}
The idea behind ADQC is simple yet surprisingly powerful, see Fig.~\ref{fig:ADQC}. The ancilla, $A$, is sequentially prepared, then entangled with a fixed interaction $E_{AR}$ to a register qubit, $R$, and then measured. The induced back-action onto the register is what ``steers'' the register to the desired state \cite{Wiseman}. An example of a universal interaction between register and ancilla is the Control-Z interaction followed by local Hadamards, $H= (X + Z)/\sqrt{2}$ with $X, Y, Z$ the Pauli matrices,
\begin{equation}
	E_{AR} = H_{A} \, H_R \, \CZ_{AR},
\end{equation}
where Control-Z is given by $\CZ = \mathbbm{1} - 2 \, |11\>\<11|$. This is reminiscent of MBQC where resource states for computation are constructed using the $\CZ$ operation \cite{RB01,BB06}.

%%%%%%%%%%%%%%%%%%
\begin{figure}[t]
   \begin{center}
	\includegraphics[width=0.38\textwidth]{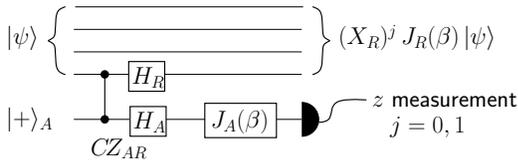}
        \caption{\label{fig:J-implement} Ancilla-driven implementation of a single qubit rotation, $J_R (\beta)$, on a register qubit $R$. The initial state of the total register can be a pure state, $| \psi \>$, or accordingly for a mixed state. The ancilla and register qubits are first coupled with $E_{AR} = H_{A} \, H_R \, \CZ_{AR}$. The rotation $J (\beta)$ is then implemented on the fully controlled ancilla and transferred to the register qubit by measuring the ancilla in the $z$-basis. The result of the measurement, $j=0,1$, determines if an $X$ correction appears on the register qubit.}
   \end{center}
\end{figure}
%%%%%%%%%%%%%%%%%%

Any arbitrary single qubit unitary can be decomposed into four rotations, $J (\beta) = H \, e^{i {\beta \over 2} X}$, as $U = e^{i \alpha} \, J(0) \, J(\beta) \, J(\gamma) \, J(\delta)$ with parameters $\beta$, $\gamma$, $\delta$ (Euler angles) and $\alpha$ (global phase) in $\mathbbm{R}$ \cite{DKP}. To implement a rotation $J_R (\beta)$ on the register in the ancilla-driven model the ancilla is first prepared in the state $|+\rangle_A$ and then coupled to the qubit $R$ via $E_{AR}$, see Fig.~\ref{fig:J-implement}. Instead of acting on the register qubit the rotation $J_A(\beta)$ is applied to the ancilla and transferred to the register qubit by measuring the ancilla in the computational $z$-basis $|j\rangle$ with $j=0, 1$. Alternatively, the ancilla can be measured immediately in the rotated basis 
$|\beta_{+} \>_A = \cos {\beta \over  2} |0\>_A + i \sin {\beta \over 2} |1\>_A$ 
and $|\beta_{-} \>_A = \cos {\beta  \over 2} |1\>_A + i \sin {\beta \over 2} |0\>_A$. The implemented operation on the register qubit is 
\begin{equation}
	{}_A\< j| J_A(\beta) \, E_{AR} \, |+\>_A  = U_R (j) \, J_{R}(\beta),
\end{equation}
with (fixed) Pauli correction $U_R (j) = (X_R)^j$ that depends on the measurement outcome $j$ of the ancilla. 
(We neglect global phases of states and unitaries as they are physically insignificant.) 
The correction can be removed by changing the ancilla measurement bases of future computational operations, cf. \cite{RB01,BB06} and \cite{KOBAA09}.

%\medskip

%%%%%%%%%%%%%%%%%%
\begin{figure}[t]
   \begin{center}
	\includegraphics[width=0.44\textwidth]{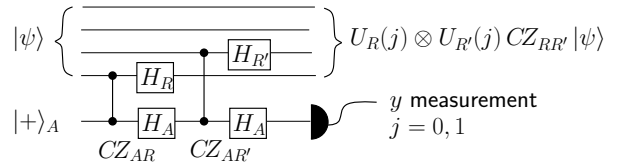}
        \caption{\label{fig:CH-parity} Ancilla-driven implementation of a Control-Z gate, $\CZ_{RR'}$, on two register qubits $R$ and $R'$. Both register qubits are coupled with $E_{AR (AR')} = H_A \, H_{R (R')}\, \CZ_{AR (AR')}$ to the ancilla which is then measured in the $y$-basis. The corrections $U_{R} (j)$ and $U_{R'}(j)$ are local and can be applied through ancilla-driven single qubit rotations.}
   \end{center}
\end{figure}
%%%%%%%%%%%%%%%%%%

Arbitrary single qubit unitaries composed with an entangling operation, such as the $\CZ$ gate, form a universal set of gates, allowing the construction of any unitary evolution. To entangle two register qubits, $R$ and $R'$, they each interact with the ancilla via the same operation $E$, see Fig.~\ref{fig:CH-parity}. A $y$-measurement of the ancilla then mediates the entangling operation between the register qubits, 
\begin{eqnarray} \label{eq:czrr}
	{}_A\< y_j| E_{AR} \, E_{AR'} \, |+\>_A =  U_{R} (j)  \otimes U_{R'}(j) \, \, \CZ_{RR'}.
\end{eqnarray}
This is a Control-Z operation up to local corrections 
$U_{R, R'} (j) = H_{R, R'} \, \left( {\mathbbm{1}_{R, R'}+ i Z_{R, R'} \over \sqrt{2}}\right) \, (Z_{R, R'})^j $ that again depend on the outcome $j$ of the ancilla measurement. $U_{R} (j)$ and $U_{R'} (j)$ can be applied explicitly as single qubit operations as described above. Thus ADQC with the interaction being fixed to $E_{AR} = H_{A} \, H_R \, \CZ_{AR}$ allows the implementation of any unitary evolution, or \emph{universal state preparation}, of the register \cite{footnote-twisted}.

\smallskip

{\it General interactions for ADQC.--}
A fundamental question in the context of new models for quantum computation is to specify \emph{all} entangling operations, $E$, that lead to universality \cite{USP,CTN}. To add structure to this question one can restrict to computations with a number of desirable properties. 
An important requirement for the working of ADQC is that no operation, including any corrections, ever needs direct implementation on the register. We therefore consider unitary, step-wise deterministic, ``tensor-commuting'' (as defined below) computations; i.e. by adapting ancilla measurement bases alone corrections on the register can be absorbed and the computation remains deterministic at every step \cite{Flow06,g-flow}.

While in MBQC with the cluster state only standard $X$, $Y$ and $Z$ corrections occur, here we allow a broader class which we entitle \emph{generalized Pauli corrections}.  This is the class of all single-qubit Hermitian unitaries $P$ which satisfy $\tr[P] = 0$. These can be parametrized as $P (a,b,c) = a X + b Y + c Z$ with $a, b, c \in \mathbbm{R}$ and $a^2+b^2+c^2=1$. 
Let us now introduce the canonical decomposition \cite{ZVSW2003,KC01} for two-qubit unitaries,
$E_{AR} =(W'_A \otimes W_R) \, D_{AR} \, (V'_A \otimes V_R)$, which separates the
essential non-local part
\begin{eqnarray}\label{e-Dformula}
  D_{AR} (\alpha_x, \alpha_y, \alpha_z) = e^{-i (\alpha_x X_A \otimes X_R + \alpha_y Y_A \otimes Y_R + \alpha_z Z_A \otimes Z_R)}
\end{eqnarray}
with $0\le\alpha_{x,y,z}\le\pi/4$ from local single qubit unitaries on the register, $V_{R}, W_{R}$, and the ancilla, $V'_{A}, W'_{A}$. To allow composition of operations to arbitrarily large computations all corrections on register qubits need to interchange with future entangling operations, $D_{AR}$, in such a way that they remain localized, i.e. the resulting correction is a tensor product between the register and ancilla. This allows the corrections on each register qubit to shift through the computational pattern and accumulate at the final step where they can be removed by a \emph{local} redefinition of the computational basis \cite{footnote-correction}. If there exist such corrections, we say that  the entangling operation, $D_{AR}$,  \emph{tensor-commutes} with the corrections \cite{footnoteflow}, i.e. 
$D_{AR} \, P_R (a,b,c) \, D^{\dag}_{AR}  = P_A (\tilde{a},\tilde{b},\tilde{c}) \otimes P_R (a',b',c')$ where $\tilde{P_A}$ and $P'_R$ are also generalized Pauli transformations (or the identity, see \cite{AABKO}).  

The key result of this \paper is that only two classes of couplings have the potential to be universal as summarized in the theorem below.

\smallskip

{\bf Theorem. \it The interactions, $E_{AR}$, between any register qubit, $R$, and the ancilla, $A$, that result in a i) unitary, step-wise deterministic evolution of the register, that ii)  tensor-commute with corrections and that iii) admit universal state preparation of the register using ADQC, are locally equivalent to (i.e. $D_{AR}$ is of the form) the \emph{Ising model} or the \emph{Heisenberg XX model} with maximal coupling strength $\alpha=\pi/4$ or $\alpha_x=\alpha_y =\pi/4$. }

\smallskip

Moreover, any generalized measurement, or POVM, on the register can be performed in the ancilla-driven model with the (repeated) help of a second ancilla that introduces extra degrees of freedom to form a suitable Neumark extension of the register's state \cite{AO08,Rez02}. This is of interest especially when measurements on the register would remove the physical qubit, such as  photon measurements. In ADQC, the destructive measurement is instead made on the ancilla and the register qubits remain for future operations. 

%\medskip 

The importance of the theorem is in the \emph{only if} part which proves for the first time the necessity of this type of building block. The two classes of interactions are equivalent to the Control-Z gate, as shown constructively in the example above, and the Control-Z+SWAP gate, where $\CZ$+SWAP $ = |00\>\<00| + |01\>\<10| + |10\>\<01| - |11\>\<11|$. Both interactions are maximally entangling \cite{maximally-entangling,KC01} and Clifford operations \cite{Clifford}. 
%Interestingly, this contrasts with the result that Clifford operations are \emph{not} universal as a set of gates \cite{Clifford}.

%\medskip

The proof of the theorem is involved and technical details can be found in
\cite{AABKO}. An outline of the arguments is given here. When the ancilla with
initial state $|+\>_A$ and a single register qubit are coupled, and the ancilla
is subsequently measured in $|\pm\>_A$ then the state $\ket{\psi}$ of the
register (and similarly for mixed states) becomes
\begin{equation} \label{eq:cp-map} \ket{\psi} \mapsto {\KO^{\pm}_R \over
    \sqrt{p^{\pm}}} \, \ket{\psi},
  %	\rho \mapsto \rho^{\pm} = { \mop{\KO}^{\pm}_R \, \rho \, \mop{\KO}^{\pm
  %     \, \dagger}_R \over \tr[ \mop{\KO}^{\pm}_R \, \rho \, \mop{\KO}^{\pm \,
  %     \dagger}_R ]},
\end{equation}
where the Kraus operators $\mop{\KO}^{\pm}_R$ are given by
\begin{equation}
  \mop{\KO}^{\pm}_R = {}_A\bra{\pm}\, E_{AR} \, \ket{+}_A
  = _A\bra{\pm_{\ta,\phi}} \,  W_R  \, D_{AR} \, V_R \, \ket{+_{\gamma,\delta}}_A,
\end{equation}
and $p^{\pm} = \tr[\KO^{\pm \, \dag}_R \KO^{\pm}_R]$ are the probabilities of
obtaining measurement outcome $+$ or $-$.  Here we substituted
$\ket{+_{\gamma,\delta}}_A = \cos {\gamma \over 2} \ket{0}_A + e^{i \delta} \sin
{\gamma \over 2} \ket{1}_A = V'_A \ket{+}_A$ and $\ket{\pm_{\ta,\phi}}_A =
W'^{\dag}_A \ket{\pm}_A$ where $\gamma, \delta, \ta$ and $\phi$ denote ancilla
parameters contained in $V'_A$ and $W'_A$. 

In a step-wise deterministic computation the Kraus operators must necessarily be proportional to unitaries and this implies that one of the $\alpha$'s, say $\alpha_z$, must vanish. Additionally, the two Kraus operations shall relate to another via a generalized Pauli, $\KO^{-}_R / \sqrt{p^{-}}= e^{i \Delta} \, P_R \, \KO^{+}_R/ \sqrt{p^{+}}$, where $\Delta$ is an unimportant global phase. Moreover, the non-local part of the interaction, $D_{AR}$, must tensor-commute with this correction $P_R$. These requirements restrict the interaction to four classes, $D_{AR} (\pi/4, \pi/4, 0), D_{AR} (0<\alpha_x< \pi/4, \pi/4, 0), D_{AR} (\pi/4, 0, 0)$ and $D_{AR} (0 < \alpha < \pi/4, 0, 0)$, each with their individual sets of acceptable ancilla parameters $\gamma, \theta, \delta, \phi$ and sets  $P(a,b,c)$ with $a,b,c$ of tensor-commuting corrections.

However, two of these classes are not sufficient for universal state preparation. Ising interactions with non-maximal interaction strength, $D_{AR} (0<\alpha < \pi/4, 0, 0)$, can be used to steer unitary, step-wise deterministic evolutions of a register qubit. Yet the ancilla parameters are so restricted that the set of implementable single qubit unitaries lies in a plane of the Bloch sphere and $D_{AR} (0<\alpha < \pi/4, 0, 0)$ (plus local unitaries) is \emph{not universal}. For Heisenberg models with non-maximal coupling strength, $D_{AR} (0<\alpha_x<\pi/4, \pi/4, 0)$, it is impossible to \emph{compose} several single qubit operations after another while preserving step-wise determinism, see \cite{AABKO} for details. This leaves only two universal classes, $D_{AR} (\pi/4, 0, 0)$ and $D_{AR} (\pi/4, \pi/4, 0)$, which are locally equivalent to $\CZ_{AR}$ and $\CZ_{AR} + \mbox{SWAP}_{AR}$. This leads to the theorem. 
 
We note that the choice of appropriate \emph{local unitaries} in $E_{AR}$ is not trivial - $\CZ$ alone can not steer all register evolution as no basis change can be achieved at the register qubit. However, we showed in the example above that together with local Hadamards that enable a basis change, the $\CZ$ interaction can be made universal. For the $\CZ$+SWAP interaction it is easy to verify that the ADQC model is identical to the one-way model \cite{RB01} and hence allows universal state preparation, as the role of ancilla and register qubits are simply swapped. 

\smallskip

{\it Physical systems suitable for ADQC.--}
ADQC is suited to many physical realizations, e.g. a register of atoms trapped in an optical lattice addressed by ancilla marker atoms~\cite{CDJWZ2004}, which interact via cold collisions to generate control-Z gates~\cite{JBCGZ1999}; similarly an array of ions in microtraps and an ancilla read-write ion that interacts by laser-induced state-dependent pushing forces~\cite{CZ2000}. Using optimized control pulses~\cite{PSTC2009} it may be possible to generate the $E_{AR}$ operation efficiently and robustly in a single step. 
We can also consider different systems for the role of register and ancilla, e.g. a cavity QED~\cite{BBGDSGKDRH2008} register, and Rydberg atoms traversing them as ancillas~\cite{GVTE2000}. While `static' qubit field states are hard to manipulate directly, `flying' Rydberg atoms are easily controlled by lasers and measured by field-ionization. We note that~\cite{GVTE2000} suggests entangling two cavity qubits by the sequential interaction of the $XX$-type with an ancilla Rydberg atom which is then measured (c.f. Eq.(\ref{eq:czrr})). However different cavity qubit rotations are achieved by varying the control pulse between the register and ancilla, while ADQC requires only a single register-ancilla operation.

Another hetero-qubit scheme uses nuclear and electron spins, e.g. an array of long-lived phosphorus donor nuclear spins in silicon as the register~\cite{Kane1998}, and electron ancillas move around the array by charge transport by adiabatic passage (CTAP)~\cite{GCHH2004}. Control and characterization of the nuclear-electron interaction is reduced to optimization of a single two-qubit unitary, simplifying the task considerably, especially for qubits subject to manufacturing variation and tolerances~\cite{SO2009}.  We can exploit spin-orbit effects to introduce anisotropies into the Heisenberg interaction~\cite{Kavokin2001,WL2002} which allow Ising-type entangling unitaries to be generated~\cite{BL2005}. Finally, superconducting qubits coupled with an effective Hamiltonian of the $XX+YY$ type via a superconducting microwave stripline~\cite{MCGKJSFSHWBDGS2007} are also good candidates for ADQC.
  
%\medskip

In summary, ADQC is a practical method of implementing any quantum channel on a quantum register \emph{without direct manipulation}, when an entangling operation $E_{AR}$ is naturally available. Imperfections in the interaction could arise in a number of ways, for example in decoherence and in the timing and finite duration of the measurements. An important project for future work is to investigate how quantum error correction techniques can be employed to deliver fault tolerance in the ADQC model and allow implementation in a yet broader class of physical systems.

The equivalence to other computational schemes makes ADQC a valuable model that shifts the question of universal resources away from resource states and their structure, and focuses instead on basic building blocks. These can be characterized systematically by requiring properties sufficient for universal computation and showing why some fail. In the future this approach can be adapted to investigate computations with relaxed properties. For instance, one might not require the computation to be step-wise deterministic, similar to the computation using computational tensor network states \cite{CTN}. We expect that these correspond to unitary Kraus operators, i.e. one of the $\alpha$'s must vanish, however the branching relation could be non-Pauli but instead any finite root of $\mathbbm{1}$. This would lead to schemes based on Ising or Heisenberg interactions with smaller coupling strength, $\alpha < \pi/4$.

\smallskip

\small 
%{\it Acknowledgements.--}
We thank Mark Hillary for insightful discussions. JA is funded by the EPSRC's
QIPIRC network in the UK. DEB thanks the National Science Foundation of
Singapore. JA and DEB acknowledge QNET support in the UK. We all thank the
Scottish Universities Physics Alliance and QUISCO.


\begin{thebibliography}{99}
\bibitem{footnote-correction} 
The final correction can also be implemented explicitly in the ancilla-driven manner. This would be the only non-deterministic instance where repeated application may be neccessary.
\end{thebibliography}
\end{document}